\documentclass[twocolumn,showpacs,amsmath,amssymb,prl]{revtex4-1}
\usepackage{graphicx}
\usepackage{dcolumn}
\usepackage{bm}

\begin{document}
\input{epsf}

\title{Observing Lense-Thirring Precession in Tidal Disruption Flares}
\author{Nicholas Stone}
 \email{nstone@cfa.harvard.edu}
 \author{Abraham Loeb}
\affiliation{Astronomy Department, Harvard University, 60 Garden St.,
Cambridge, MA 02138, USA}

\begin{abstract}

When a star is tidally disrupted by a supermassive black hole (SMBH),
the streams of liberated gas form an accretion disk after their return
to pericenter.  We demonstrate that Lense-Thirring
precession in the spacetime around a rotating SMBH can produce
significant time evolution of the disk angular momentum vector, due to both the periodic precession of the disk and the nonperiodic, differential precession of the bound debris streams.  Jet precession and periodic modulation of disk luminosity are possible consequences.
The persistence of the jetted X-ray emission in the Swift
J164449.3+573451 flare suggests that the jet axis was aligned with the
spin axis of the SMBH during this event.

\end{abstract}

\pacs{98.62.Js, 98.62.Mw, 98.62.Nx}

\date{\today}
\maketitle

\paragraph*{Introduction.}
The tidal disruption of a star by a supermassive black hole offers a
unique opportunity to probe the nuclei of otherwise quiescent
galaxies.  However, the small number of candidate tidal disruption events
(TDEs) makes it difficult to resolve theoretical uncertainties concerning their rates \cite{Donley2002, WM2004, MP2004, Perets2007, Gezari2008, vanVelzen2010}, super-Eddington
accretion phase \cite{LU1997, SQ2009, SQ2011}, and the period during which dissipation in shocks
allows an accretion disk to form \cite{Kochanek1994, Ulmer1999}.

An additional outstanding question about TDEs is whether or not they
produce jets, as observed in many other accreting black hole systems.
The past year has seen both the first theoretical models for
TDE-associated jets \cite{GM2011, vanVelzen2011b}
and the discovery by the Swift satellite of an intense, transient
gamma- and X-ray flare from a galactic nucleus at $z\approx 0.35$
\cite{Levan2011}.  This flare has been explained by multiple
authors \cite{Zauderer2011, Bloom2011, Burrows2011}
as jet emission from a TDE aligned with our line of sight (although
alternate hypotheses exist \cite{QK2011}).  A second possible TDE-associated jet was also recently observed \cite{Cenko2011}.

If such jet emission is common, then TDEs provide a unique probe of the
physics of accretion and jet production in the vicinity of
distant black holes' horizons.  Specifically, it is unknown at present whether
jets will align with the black hole spin vector, the disk angular
momentum vector, or some other component of the magnetic field
geometry \cite{Fragile2008}.  In most black hole accretion environments
these directions are parallel, but the transient disk of a TDE will
generally have some tilt with respect to the SMBH equatorial plane.
In this {\it Letter} we demonstrate that if jets from tilted TDE accretion
disks align with the disk normal vector, they will generally be
expected to precess, often by observable amounts.  Even absent the
existence of a disk-aligned jet, or any jet at all, general
relativistic (GR) effects will precess TDE disks with potentially
observable consequences.

\begin{figure}
\includegraphics[height=44mm]{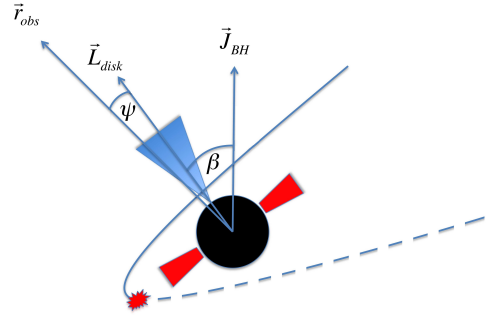}
\caption{Geometry of the tidal disruption of a star by a spinning
SMBH.  Following disruption of the star near
its pericenter passage, an accretion disk will form in the star's
orbital plane.  As the disk precesses, the angle $\beta$ between the
SMBH spin vector $\vec{J}_{\rm BH}$ and the disk angular momentum
vector $\vec{L}_{\rm disk}$ stays constant, but an associated jet may
move relative to the observer's line of sight $\vec{r}_{\rm obs}$.} \label{5}
\end{figure}

\paragraph*{Spin evolution of a tilted disk.}
Stars of mass $M_*$ and radius $R_*$ that pass within a radius
\begin{equation}
R_{\rm t}=R_*(M_{\rm BH}/M_*)^{1/3}
\end{equation}
of a black hole of mass $M_{\rm BH}$ will be tidally disrupted, with
half their mass immediately unbound from the black hole \cite{Rees1988}.
For black holes of mass $M_{\rm BH}\gtrsim 10^8M_{\odot}$, the tidal
radius $R_{\rm t}$ is inside the Schwarzschild radius $R_{\rm S}$ and
stars are swallowed whole rather than disrupted.  The bound debris
rapidly expands and cools so that its pressure is negligible
and the approximation of geodesic motion is accurate \cite{Kochanek1994}.
The most tightly bound debris stream of a star disrupted at radius
$R_{\rm P}$ returns in a time
\begin{equation}
t_{\rm fall}\approx 50~M_6^{5/2}r_{\rm p}^3 r_*^{-3/2}~{\rm s},
\end{equation}
where $M_6=M_{\rm BH}/10^6M_{\odot}$, $r_*=R_*/R_{\odot}$ and $r_{\rm
p}=R_{\rm p}/R_{\rm S}$ \cite{SQ2009}, although $t_{\rm fall}$ depends on the stellar density profile and can be evaluated more precisely by numerical simulations  \cite{Laguna1993}.  After a small
multiple of this time, stream-stream collisions circularize the
returning gas and allow an accretion disk to form.  In general, this
transient accretion disk will not lie in the black hole equatorial
plane.

An accretion disk inclined out of the equatorial plane of a spinning
black hole by an angle $\beta$ (assumed to equal the inclination of the stellar orbit before disruption, $\beta_*$ - see Fig. \ref{5}) will be subject to Lense-Thirring torques with a strong
radial dependence.  For a thin disk \cite{KP1985}, it is
expected that the Bardeen-Petterson effect \cite{BP1975, PP1983} will induce a warp in the disk structure.  However, for the thicker disks expected in many TDEs \cite{Ulmer1999, SQ2009}, simulations combining GR and magnetohydrodynamic effects (GRMHD) have shown that the disk
precesses as a solid body rotator \cite{Fragile2007, DF2011}.  Such an accretion disk will precess with a period $T_{\rm prec}=2\pi {\rm sin}\beta(J/\tau)$,
where $J$ is  total angular momentum and $\tau$ is integrated torque.  A notable feature of this formula is that $T_{\rm
prec}$ is independent of many disk model parameters, and depends only
on the dimensionless radial surface density profile.

The simulations mentioned above considered disks with a roughly constant
surface density.  For surface densities of the form $\Sigma=\Sigma_{\rm i} (R/R_{\rm
i})^{-\zeta}$, the precession timescale is \cite{Fragile2007}
\begin{equation}
T_{\rm prec} = \frac{8\pi GM_{\rm BH} (1+2\zeta)}{c^3(5-2\zeta)} \frac{r_{\rm o}^{5/2-\zeta}r_{\rm i}^{1/2+\zeta}(1-(r_{\rm i}/r_{\rm o})^{5/2-\zeta})}{a(1-(r_{\rm i}/r_{\rm o})^{1/2+\zeta})}.
\end{equation}
Here the disk inner ($R_{\rm i}$) and outer ($R_{\rm o}$) edges have been normalized to units of Schwarzschild radii ($r_{\rm i}=R_{\rm i}/R_{\rm S}$, $r_{\rm o}=R_{\rm o}/R_{\rm S}$).
The variable $a$ is the dimensionless black hole spin
parameter, with values between 0 and 1.

Whether or not the disks associated with tidal disruption flares
approximately follow a surface density profile $\Sigma=\Sigma_{\rm i}
(R/R_{\rm i})^{-\zeta}$ is unclear.  Ref. \cite{SQ2009} presented a
slim disk model for TDE accretion flows, with height $H$ given by:
\begin{equation}
\frac{H}{R}=\frac{3f}{4}\frac{10\dot{M}}{\dot{M}_{\rm Edd}}\frac{R_{\rm S}}{R}K^{-1},  \label{1}
\end{equation}
where the function $K$ is defined as
\begin{equation}
K=\frac{1}{2} + \sqrt{\frac{1}{4} + \frac{3f}{2}\left(\frac{10\dot{M}}{\dot{M}_{\rm Edd}}\right)^2 \left(\frac{R_{\rm S}}{R} \right)^2}. 
\end{equation}
Here $f=1-(R_{\rm i}/R)^{1/2}$.  $\dot{M}/\dot{M}_{\rm Edd}$ is the ratio of the mass
accretion rate to the Eddington rate assuming 10\% accretion efficiency.  

However, this model is not suitable for use in calculating $T_{\rm
prec}$, as the zero-torque boundary condition used to calculate $f$
leads to an unphysical singularity in $\Sigma$ at $R_{\rm i}$ ($\Sigma \propto R^{3/2}K^2/f$).  A
different, numerical model was recently presented in
Ref. \cite{MdFP2011}, in which axisymmetric disk equations were evolved
with a time-dependent rate of mass input at the pericenter of
disruption.  This model led to a shallow decline of $\Sigma$ with
decreasing $r$ after the arrival of the inner edge of the accretion
flow at the innermost stable circular orbit.  Motivated by Refs. \cite{SQ2009, MdFP2011}, we consider $\zeta=-3/2, 0, 1$ in this paper.  $T_{\rm prec}$ increases by a factor $\approx 7$ when going from the $\zeta=1$ to the $\zeta=-3/2$ model.

The framework we followed is based on two underlying assumptions: {\it
(i)} a coherent accretion flow exists; and {\it (ii)} the flow is not
susceptible to Bardeen-Petterson warps ($H/R \gtrsim \alpha$, where $\alpha$ is the dimensionless disk viscosity parameter). Assumption {\it (i)} is only valid after a time $t_{\rm circ}\approx n_{\rm orb}t_{\rm fall}$, where $n_{\rm orb}$ is the number of orbits
required to circularize the most tightly bound debris streams \cite{Ulmer1999}.  A value of $n_{\rm orb}\sim 1$--$10$ is often assumed
in the TDE literature, but this quantity is poorly constrained and
could be higher for large $a$ and $\beta_*$, where
Lense-Thirring precession can delay the
stream-stream collisions necessary for disk formation \cite{Kochanek1994}.
At later times, assumption {\it (ii)} will break down, as 
$\dot{M}/\dot{M}_{\rm Edd}$ declines and the disk becomes
geometrically thinner.  

\begin{figure} 
\includegraphics[width=84mm]{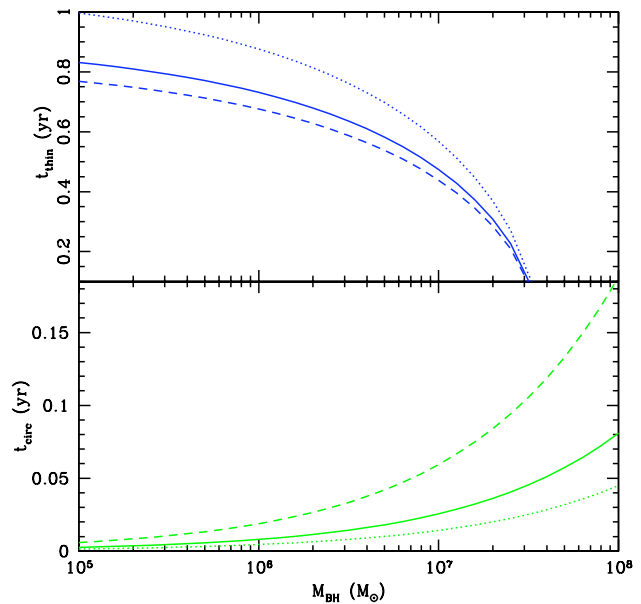}
\caption{Timescales for avoiding Bardeen-Peterson warping $t_{\rm thin}$ (blue, top panel) and for establishing an accretion disk $t_{\rm circ}$ (green, bottom panel) as functions of the black hole mass $M_{\rm BH}$.  Dotted lines
correspond to stars with a mass of $2M_{\odot}$, solid lines to
$1M_{\odot}$ and dashed lines to $0.5M_{\odot}$ (with a stellar mass-radius relationship adopted from Ref. \cite{KW}, p. 208).  We take $n_{\rm orb}=3$ and $R_{\rm p}=0.5R_{\rm t}$, and conservatively plot $t_{\rm thin}$ for the outer edge of the disk, assuming $R_{\rm o}=2R_{\rm p}$.\label{2}
}
\end{figure}

Adopting Eq. (\ref{1}) for convenience, $H/R$ will fall below $\alpha$ after a time
\begin{eqnarray}
t_{\rm thin}=t_{\rm fall}\left( \frac{5}{2} \frac{f}{X} \frac{M_*/t_{\rm fall}}{\dot{M}_{\rm Edd}} \frac{R_{\rm S}}{R} \right)^{3/5}&\\
\approx 0.3~M_6^{2/5} r_{\rm p}^{6/5} m_*^{3/5} r_*^{-3/5} \left(\frac{f}{X_{-1}} \frac{R_{\rm S}}{R} \right)^{3/5}~{\rm yr}\notag,
\end{eqnarray}
where the function $X\sim\alpha$ and is $X=\alpha/(1-8\alpha^2/3f)$.
Also note that $X_{-1}=X/0.1$.  TDE
disks will precess as solid body rotators during the time range
$t_{\rm circ}<t<t_{\rm thin}$ as illustrated in Fig.
\ref{2}, which shows that for $M_{\rm BH} \lesssim 10^7 M_{\odot}$ (and any
realistic $R_{\rm p})$, solid body precession will occur for $\lesssim
1~{\rm yr}$.

\paragraph*{Angular momenta of returning debris streams.}
The evolution of the debris streams prior to their first return to
pericenter has been studied in detail by Ref. \cite{Kochanek1994}.  The orbits of these streams, if
non-equatorial, lack a constant orbital plane due to Lense-Thirring
torques.  The accretion disk is therefore fed by a supply of new gas
with time-dependent angular momentum, which in turn evolves the
direction of $\vec{L}_{\rm disk}$.  In contrast to direct precession
of the accretion disk, we call this effect ``differential stream
precession,'' or DSP.  Although we will compute numerical general
relativistic solutions for the DSP, we can gain valuable intution from
a simpler, lowest order estimate in the post-Newtonian limit.

The angle by which the angular momentum vector of a debris stream will
precess during an orbit of period $T$ will be $\phi_{\rm orb}(T)\approx \Delta\Omega {\rm sin}(\beta)=
2\pi {\rm sin}(\beta) (T/t_{\rm LT})$, where $\Delta\Omega$ is the nodal precession and
\begin{equation}
t_{\rm LT} = \frac{T}{2a}\left(\frac{c^2 A (1-e^2)}{GM_{\rm BH}}\right)^{3/2}
\end{equation}
is the Lense-Thirring precession period \cite{Merritt2010} for a gas stream of semimajor axis $A$ and eccentricity $e$.  Defining $\Delta\phi_{\rm orb}=\phi_{\rm orb}(T)-\phi_{\rm orb}(\infty)$ as a measure of the DSP,
\begin{equation}
\Delta\phi_{\rm orb}= 4\pi a {\rm sin}(\beta)(2r_{\rm p})^{3/2}((1+e)^{-3/2}-2^{-3/2}), \label{preceq}
\end{equation}
which Taylor expands in the late-time, $R_{\rm p}/A \ll 1$ limit to $\Delta\phi_{\rm orb}\approx 1.7~{\rm sin}(\beta)ar_{\rm p}^{-5/2}r_*M_6^{-1}(t/t_{\rm fall})^{-2/3}.$

Although Eq. (\ref{preceq}) is not exact, it provides a valuable insight:
the DSP is largest for low-mass, rapidly spinning SMBHs that
disrupt stars with deeply plunging, inclined initial orbits.  At early times the disk viscous timescale $t_{\rm visc} \lesssim t_{\rm fall}$ \cite{SQ2009} so Eq. (\ref{preceq}) approximates the angular evolution of $\vec{L}_{\rm disk}$.  We do
not expect $\Delta\phi_{\rm orb}>1^\circ$ after the establishment
of a steady accretion flow ($t>3t_{\rm fall}$) for any TDEs with solar-type stars and $M_{6}\gtrsim 2$, although these constraints relax for stars with
$r_*>1$.

\begin{figure}
\includegraphics[width=84mm]{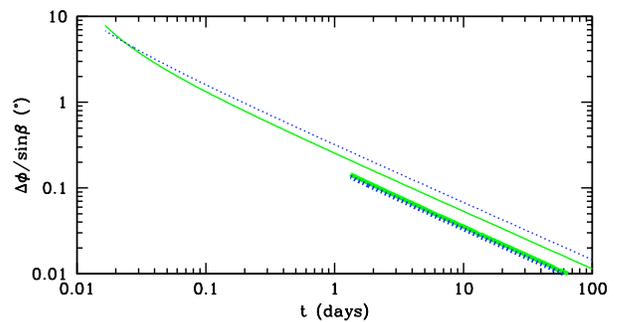}
\caption{The angular shift $\Delta\phi_{\rm orb}$.  The thick curves illustrate the disruption of a solar-type star with $M_{\rm BH}=10^6M_{\odot}$, $a=0.8$, and $r_{\rm p}=13$; the thin curves are the same but with $r_{\rm p}=3$.  The blue dotted lines are Eq. (\ref{preceq}), while the green solid lines are numerical geodesic solutions.  The curves do not extend prior to $t=t_{\rm fall}$, and are normalized by ${\rm sin}\beta$.} \label{3}
\end{figure}

To obtain an exact solution for the time evolution of angular
momentum in the returning debris streams, a GR calculation is needed.  We numerically integrate the Kerr geodesic equations following the formalism of Ref. \cite{DH2004}.  We assume a flat distribution of debris mass with specific Newtonian energy $E$, a spread in that energy of $3GM_{\rm BH}R_*/R_{\rm p}^2$ \cite{SQ2009}, and obtain constants of integration for each debris stream by transforming the initial conditions $\{E, R_{\rm p}, \beta\}$ to $\{E_{\rm GR}, L_{\rm z}, Q\}$ ($E_{\rm GR}, L_{\rm z}, Q$ are specific energy, z-component angular momentum, and Carter's constant for Kerr metric test particles).  Good agreement with Eq. (\ref{preceq}) is shown in Fig. \ref{3}.

\paragraph*{Observational implications.}
We have shown that the Lense-Thirring effect will cause the direction
of a TDE disk's angular momentum vector to evolve in time.  Direct precession of the
accretion disk is the dominant effect, but in some cases DSP can cause a significant non-periodic
evolution in $\vec{L}_{\rm disk}$.  The precession of the accretion disk will modulate the observed disk
luminosity at least by a factor of ${\rm cos}(\psi)$, and lead to
periodic pulsations of the associated transient quasar.  This periodic
modulation could in principle be extracted from the Fourier
decomposition of a TDE lightcurve, but perhaps a more promising avenue
for detection lies in the fraction of events for which the disks will
precess into an edge-on phase.  This could reduce the observed disk
flux by $\sim 2$ orders of magnitude while simultaneously reddening
the peak emission frequency \cite{Ulmer1999}.  Even in the absence of jet
emission, observations of a ``blinking'' TDE flare could provide
strong evidence of precession and allow both $a$ and the
disruption parameters to be constrained.  

The most exciting possible consequence, however, is precession of jets
associated with TDE disks.  If we assume that relativistic jets in
tilted accretion systems align with $\vec{L}_{\rm disk}$, narrow jets will precess out of the observer's line of sight
in a small fraction of $T_{\rm prec}$.  Continuous observation of a jet for a relatively short period of time,
$t_{\rm obs}$, allows very strong constraints to be placed on
combinations of $a$ and disruption parameters such as
$r_{\rm p}$ and $\beta_*$ (assuming still that $\beta=\beta_*$).  Alternatively,
repeated observation of TDE-associated jets could serve as evidence
that jets align with $\vec{J}_{\rm BH}$ or an aspect of the
magnetic field geometry, provided that sufficient non-precession is observed.  We note that the DSP, though generally subdominant, can in some cases cause very rapid precession (up to $\sim 0.1^{\circ}/{\rm min}$) at the onset of the flare (Fig. \ref{3}).  If an associated jet is aligned with $\vec{L}_{\rm disk}$, this will lead to a brief, nonrepeating transient which could fake an unusually long gamma ray burst provided $\theta_{\rm jet} \lesssim 1^{\circ}$.

To provide a concrete example of the above considerations, we consider
the tidal disruption candidate Swift J164449.3+573451,  for which Ref. \cite{Zauderer2011} inferred the following relevant
disruption parameters: $M_{\rm BH} \sim 10^5-10^6M_{\odot}$, $R_{\rm
p} \approx 13R_{\rm S}M_6^{-5/6}$, and $\theta_{\rm jet} \sim
10^{-1.5}$ ($\theta_{\rm jet}$ is estimated from both comparing the theoretical TDE rate to the observed rate of jets over the period of the Swift mission, and the Eddington limit of the SMBH).

\begin{figure}
\includegraphics[width=84mm]{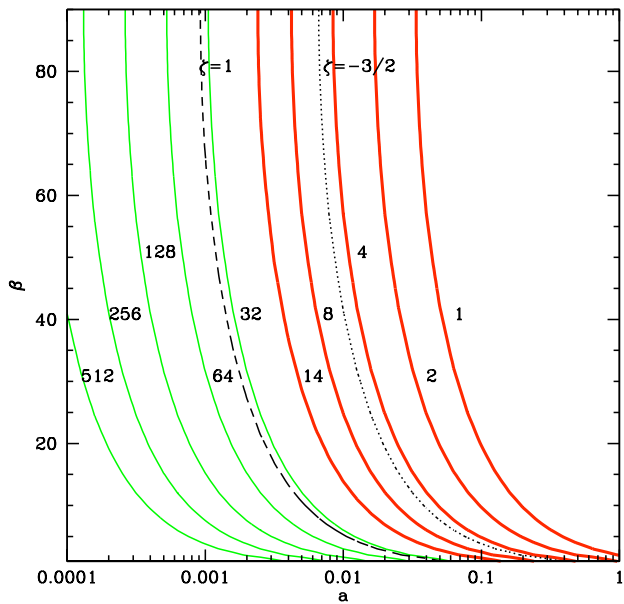}
\caption{Regions of $a$-$\beta$ parameter space that can
be excluded by continuous observations of a TDE jet with the inferred
parameters in Ref.  \cite{Zauderer2011} and $\zeta=0$.  The solid curves show contours of
constant $t_{\rm obs}=T_{\rm prec}\times2(\theta_{\rm jet}/10^{-1.5})/(2\pi{\rm sin}\beta)$: the
maximum number of days it would take for a jet initially in the
observers' line of sight to precess off-axis, with the jet opening angle normalized to $10^{-1.5}$.  We take $R_{\rm o}=2R_{\rm p}$ and $R_{\rm i}=3R_{\rm S}$.  Regions of parameter space to the right of the thick red
contours can be excluded for the Swift TDE jet, which
exhibited bright X-ray emission for over two weeks.  The 14 day contours for $\zeta=-3/2$ and $\zeta=1$ are shown with black dotted and dashed lines, respectively.  The effect of the DSP is negligible for these parameters, and neglected here.}
\end{figure}

Figure 4 shows the resulting constraints on the joint $a$-$\beta$ parameter space of this TDE if we take $M_{6}=1$.  Since the bright X-ray
emission from Swift J164449.3+573451 persisted for over two weeks, at
least one of the following statements must be true: {\it (i)} the
value of $a$ is extremely low, $\lesssim 10^{-2}$ ($10^{-1}$ if $\zeta=-3/2$); {\it
(ii)} the initial orbit of the disrupted star was tightly aligned to
within $\sim \theta_{\rm jet}$ of the black hole equatorial plane; or {\it
(iii)} the jet emission was not aligned with the disk spin axis.  The first possibility would represent an unusually low value of black hole spin and could be excluded if the Blandford-Znajek mechanism was responsible for jet launching \cite{LZ2011}, while {\it (ii)} requires that 
there will be a larger abundance of somewhat shorter events.  Since
such flares are not frequently observed, the persistent X-ray emission
in Swift J164449.3+573451 suggests that its jet was aligned with the
steady spin axis of the black hole rather than with its precessing
disk. Future GRMHD simulations can test this
inference from first principles.  The detection of additional
TDE-associated jets in future surveys 
would test the statistical robustness of this conclusion.

\bigskip
\paragraph*{Acknowledgments.}
This work was supported in part by NSF grant AST-0907890 and NASA
grants NNX08AL43G and NNA09DB30A.  We thank Ashley Zauderer and both anonymous referees for useful comments.

\bibliography{draft4}

\end{document}